\documentclass[12pt]{article}% scrbook, scrreprt, scrlettr, scrartcl
\pdfoutput=1
\usepackage[top=0.7in,bottom=1in,left=1.1in,right=1.1in,includehead]{geometry}
\usepackage{epsfig,amsmath,amsfonts}
\usepackage{hyperref}

\def\a{\alpha} \def\b{\beta} \def\g{\gamma} \def\d{\delta} \def\e{\epsilon}
  \def\h{\eta} \def\q{\theta}
    \def\m{\mu}
\def\n{\nu}    \def\r{\rho}
 \def\s{\sigma}   
   \def\w{\omega}

\def\G{\Gamma} 
 
   \def\Q{\Theta}
    
 \def\X{\Xi}  
 \def\S{\Sigma}  
   \def\W{\Omega}

\def\fr{\frac}  \def\dt{\partial}

\def\ph{\phantom}
\def\mc{\mathcal}

\def\Tr{\mbox{Tr}}

\newcommand{\beq}[1]{\begin{eqnarray}\label{#1}}
\newcommand\eeq {\end{eqnarray}}
\newcommand\bqa {\begin{eqnarray}}
\newcommand\eqa {\end{eqnarray}}

\newcommand{\bear}{\begin{array}}
\newcommand{\enar}{\end{array}}

\newcommand{\br}[2]{\bar{#1}\bar{#2}}

\def\beq{\begin{equation}}
\def\eeq{\end{equation}}
\def\bea{\begin{eqnarray}}
\def\eea{\end{eqnarray}}
\def\ba{\begin{align}}
\def\ea{\end{align}}
\def\bsp{\begin{split}}
\def\esp{\end{split}}

\begin{document}

\begin{titlepage}

\vfill
\begin{flushright}
QMUL-PH-12-21
\end{flushright}

\vfill

\begin{center}
   \baselineskip=16pt
   {\Large \bf Gauged supergravities in 5 and 6 dimensions \\ from generalised Scherk--Schwarz reductions.}
   \vskip 2cm
    Edvard T. Musaev\footnote{\tt e.musaev@qmul.ac.uk}
       \vskip .6cm
             \begin{small}
                          {\it Queen Mary University of London, Centre for Research in String Theory, \\
             School of Physics, Mile End Road, London, E1 4NS, UK} \\ 
\end{small}
\end{center}

\vfill 
\begin{center} 
\textbf{Abstract}
\end{center} 
\begin{quote}
It is shown that the Scherk--Schwarz reduction of M--theory in the Berman--Perry duality invariant formalism to 6 and 5 dimensions reproduces the  known structures of gauged supergravities that are normally associated to non--geometric compactifications. The local symmetries defined by the generalised Lie derivative reduce to gauge transformations that exactly match those given by the embedding tensor of gauged supergravity. 
\end{quote} 
\vfill
\setcounter{footnote}{0}
\end{titlepage}

\tableofcontents

\section{Introduction}

The low energy limit of M-theory is eleven dimensional maximal supergravity. When dimensionally reduced this theory possesses various duality symmetries depending on the dimension of the internal space and its geometry. Toroidal compactifications of maximal supergravity from 11 to $11-n$ dimensions lead to maximal supergravities with $E_{n(n)}$ as a global duality group and abelian $U(1)^{n_{V}}$ as a local group where $n_V$ stands for the number of vector fields. These theories have been studied extensively in the past 30 years and after a great deal of work \cite{deWit:1981eq, Nicolai:2000sc, deWit:2004nw, Bergshoeff:2007ef, deWit:2008ta, LeDiffon:2008sh} it was concluded that the so-called gaugings are  the only consistent supersymmetric deformations of these theories. 
However, in general not all gaugings of supergravities can be found as dimensional reductions of eleven dimensional supergravity. The so-called \emph{non-geometric} gaugings do not correspond to any choice of internal manifold (for a review of gauged supergravities see \cite{Samtleben:2008pe}).

A similar situation occurs when one considers backgrounds of string or M-theory which allow U-duality transformations as proper transition functions. Such backgrounds are the so-called U-folds \cite{Kumar:1996zx} and cannot be described in terms of ordinary manifolds and Riemannian geometry \cite{Andriot:2011uh, Chatzistavrakidis:2012qj, Dibitetto:2012rk}. The dimensional reduction on these non-geometric backgrounds was extensively explored for years, see for instance \cite{Flournoy:2004vn, Shelton:2005cf, Hull:2005hk, Dabholkar:2005ve, Hull:2006tp, Wecht:2007wu, Grana:2008yw} and more recently \cite{Andriot:2012an, Dibitetto:2012ia, Dibitetto:2012rk}.

The recently developed framework of the extended geometry where the space is augmented describes both geometric and non-geometric backgrounds on equal footing. This is done by introducing the so-called dual coordinates that correspond to winding modes of strings or membranes. In the case of T-duality the duality group is $O(d,d)$ and the space is just doubled \cite{Hull:2004in,Hull:2006va} giving the name of Double Field Theory \cite{Hull:2009mi, Hohm:2010jy, Hohm:2010pp, Hohm:2010xe, Hohm:2011nu}. In M-theory the situation is a little bit more subtle since the duality group jumps with dimension \cite{Hull:2007zu, Grana:2009im, Hillmann:2009ci, Aldazabal:2010ef, Berman:2010is, Berman:2011pe, Berman:2011cg, Berman:2011jh, Berman:2011kg}. In dimension $d$ one has the group $E_{d}$ as a duality group with the identification $E_4\equiv SL(5)$ and $E_5\equiv SO(5,5)$. The dimensions $d>8$ are not fully understood since they correspond to the infinite-dimensional Kac-Moody algebras $E_9$, $E_{10}$ and $E_{11}$ \cite{Riccioni:2007ni}. The U-duality covariant metric of the M-theoretic extended geometry was described in \cite{Hull:2007zu, Pacheco:2008ps} and its applications were considered in \cite{Malek:2012pw}. The geometry of the extended space in this formalism is still a mystery although certain progress in understanding the geometry of double field theory has been made  \cite{Jeon:2010rw, Jeon:2011cn,Hohm:2012gk,Hohm:2011si,Hohm:2012mf}. One cannot consistently construct a supersymmetric version of the extended geometry formalism in an ordinary way because of the dimensionality of the extended space; the supersymmetric extension of the approach based on Hitchin's generalised geometry \cite{Hitchin:2004ut, Gualtieri:2003dx, Hitchin:2005in}, where one only extends the tangent space, has recently been described \cite{Coimbra:2011nw, Coimbra:2012af}. For double field theory, significant progress has been made in  \cite{Jeon:2012hp, Jeon:2012kd, Jeon:2011sq, Jeon:2011vx, Hohm:2011nu} where one uses projectors on appropriate subspaces.

The idea of dimensionally reducing on a non-geometric background becomes very natural in the context of extended geometry. Non-geometry becomes described in geometric terms from the point of view of the extended space. The relationship between the extended geometry and gauged supergravities has been explored in \cite{Aldazabal:2011nj, Geissbuhler:2011mx, Grana:2012rr, Berman:2012uy} and involves the Scherk-Schwarz reduction \cite{Scherk:1979zr} of the extended space.

An important feature of the extended geometry is a constraint called the physical section condition or strong constraint. This condition reduces the number of dimensions to only physical ones and is required for consistency of the algebra of local transformations. In particular, it is needed to make the effective potential invariant under local gauge transformations. In general, the consistency of the theory requires all dynamical fields and their products to obey the section condition. This in turn means that the extended geometry approach is just a rewriting of the supergravity in terms of duality covariant variables. 

As it was shown in \cite{Aldazabal:2011nj, Grana:2012rr} for backgrounds that satisfy an appropriate Scherk-Schwarz anzatz this constraint can be relaxed to a weaker one. The local transformations generated by a Dorfmann or generalised Lie derivative reduce to gauge transformation. The gauge group generators are written explicitly in terms of the Scherk-Schwarz twist matrices and are identified with the embedding tensor of gauged supergravity. The closure of the algebra and the gauge invariance condition on the structure constants give the quadratic constraint that replaces the section condition.

In this paper we work with the global duality groups $SO(5,5)$ and $E_{6(6)}$ which correspond to the number of dualised coordinates $d=5$ and $d=6$ respectively. Scherk-Schwarz reduction of the extended space leads to the gauge group generators of $D=6$ and $D=5$ maximal gauged supergravities with $D=11-d$. Finally, the effective potential of the extended geometry that involves the generalised metric and its derivatives becomes the scalar potential of the corresponding maximal gauged supergravity. We show that the effective potential can be written only in terms of the gauge group generators and it is invariant under gauge transformations.  In what follows the number of compact directions is always denoted by $d$ while the number of non-compact directions is denoted by $D$.

This paper is organized as follows. In section 2, we review the Berman-Perry formalism for M-theory its local symmetries and physical condition. In section 3, we look at the Scherk-Schwarz reduction of the extended space and its relation to the embedding tensor. In section 4, we briefly review five and six dimensional maximal gauged supergravities their algebraic structure and scalar potentials. Notations for indices, coset representatives and invariant tensors are mainly introduced in this section. Finally in sections 5 and 6 we derive the algebraic structure of the Scherk-Schwarz reduced extended geometry, perform the reduction of the effective potential and identify the resulting potential. The short appendix with some details of calculations and properties of the corresponding algebras is included at the end of the paper.

\section{Extended geometry}

The duality manifest formalism  is based on the notion of an extended space. This space unifies ordinary coordinates and the so-called dual coordinates that correspond to winding modes. Such an approach allows to put both translational and winding degrees of freedom of extended objects on the same footing.

The invariant dynamics is constructed using the generalised metric that is given in terms of the background supergravity fields. Each field and each brane or string charge have corresponding coordinates: for the metric we have ordinary coordinates $x^\mu$, the NS-NS Kalb-Ramond field $B_{\m\n}$ that couples to a string leads to $d$ dual coordinates $\tilde{x}_\mu$ if U-duality acts in the $d$-dimensional space, the 3-form field of M-theory $C_{\m\n\r}$ couples to a membrane and corresponds to a dual coordinate $y_{\m\n}$. This leads to U-duality covariant formulation of a theory. 

While in the case of type II string theory the T-duality group is $O(d,d)$ for any dimension $d$, the M-theory case is more tricky since the duality group is different in different dimensions.
\begin{center}
 \begin{tabular}{|c|c|c|c|c|}
\hline
 d&  Global duality group & Local duality group & $\mc{R}_V$\\
\hline 
  4 & $SL(5)$ 	& $SO(5)$			& $\bf{10}$\\
  5 & $SO(5,5)$	& $SO(5)\times SO(5)$ & $\bf{16}$	\\
  6 & $E_6$	& $USp(8)$		& $\bf{27}$\\
  7 & $E_7$	& $Sp(8)$	& $\bf{56}$		\\
  8 & $E_8$	& $SO(16)$	& $\bf{248}$	\\
\hline  
 \end{tabular}
\end{center}
There are two main reasons for such a behaviour. The first is that fundamental objects of M-theory are known to be 2- and 5-branes. That is for $d<5$ the 5-branes do not play a role and their winding modes do not contribute. The other reason is that the number of winding modes for a 2-dimensional object in $d$ compact dimensions is given by $d(d-1)/2$. Compare with the stringy case where the number of dual coordinates is always $d$ for $d$ compact dimensions. Under the action of the duality group vector multiplets in the supergravity with $d$ compact dimensions transform as the representation $\mc{R}_V$ which is listed in the table above.

A key geometric structure  on the extended space is the generalised Lie derivative. It is defined by the Dorfmann bracket that is a natural product of elements of the generalised tangent bundle \cite{Hitchin:2004ut, Gualtieri:2003dx}.

Consider an extended space $\mc{S}$ with coordinates $X^M$ which emerges as an extension of the ordinary $d$-dimensional space and is labelled by the index $M$ that runs from 1 to $n_V=\mbox{dim}\mc{R}_V$.  We will ignore the other space-time coordinates that do not have duals.

The transformation of tensors that is consistent with the Courant bracket $[\;,\;]_C$ is defined by the generalised Lie derivative \cite{Grana:2008yw}
\begin{equation}
\label{Lie}
\begin{split}
&\d_\S Q^M=(\mc{L}_\S Q)^M=(L_\S Q)^M+Y^{MN}_{KL}\dt_N\S^K Q^L, \\
&[Q_1,Q_2]_C=\fr12\left(\mc{L}_{Q_1}Q_2-\mc{L}_{Q_2}Q_1\right).
\end{split}
\end{equation}
Here both the transformation parameter $\S^M$ and the vector $Q^M$ are functions of the extended coordinate $X^M$ and the tensor $Y^{MN}_{KL}$ is an invariant tensor of the corresponding U-duality group and it is basically a projector \cite{Coimbra:2011ky}:
\begin{equation}
\label{Y}
\begin{array}{rcl}
O(d,d)_{strings}: & \quad & Y^{MN}_{PQ}  = \eta^{MN} \eta_{PQ},, \vspace{0.2cm} \\
SL(5):  & \quad & Y^{MN}_{PQ}= \e^{\a MN}\e_{\a PQ},   \\[0.2cm]
SO(5,5): &\quad & Y^{MN}_{PQ}  = \frac{1}{2} (\G^i)^{MN} (\G_i)_{PQ} \ ,  \\[0.2cm]
E_{6(6)}: &\quad & Y^{MN}_{PQ}  = 10 d^{MN R} d_{PQR} \ ,  \\ [0.2cm]
E_{7(7)}: &\quad & Y^{MN}_{PQ}  = 12 c^{MN}{}_{PQ} + \delta^{(M}_P \delta^{N)}_Q + \frac{1}{2} \e^{MN} \e_{PQ } \ . 
\end{array}
\end{equation}
Here the index $\a$ runs from 1 to 5 labelling the representation $\bf{5}$ of $SL(5)$ and the index $i$ labels the 10-dimensional vector representation of $SO(5,5)$. The invariant metric on $O(d,d)$ is denoted by $\h_{MN}$, $\e_{\a MN}=\e_{\a,\b\g,\d\e}$ is the $SL(5)$ alternating tensor, $SO(5,5)$ gamma-matrices $\G^{iMN}$ are $16\times16$ gamma-matrices in Majorana-Weyl representation, the tensors $d_{MNK}$ and $c^{MN}{}_{KL}$ are symmetric invariant tensors of $E_6$ and $E_7$ respectively.

The generalised Lie derivative $\mc{L}_\S$ is a deformation of the ordinary Lie derivative ${L}_\S$. The deformation is given by the invariant tensor $Y_{KL}^{MN}$ that is subject to various important relations that will be used later \cite{Berman:2012vc}
\begin{equation}
\begin{split}
\label{rel}
&Y^{(MN}_{KL}Y^{L)R}_{PQ}-Y^{(MN}_{PQ}\d^{R)}_{K}=0 \mbox{ , for $d\leq5$},\\
&Y^{MN}_{KL}=-\a_d P_{K}{}^{M}{}_{L}{}^{N}+\b_d\d^M_K\d^N_L+\d^M_L\d^N_K,\\
&Y^{MA}_{KB}Y^{BN}_{AL}=(2-\a_d)Y^{MN}_{KL}+(D\b_d+\a_d)\b_d\d^M_K\d^N_L+(\a_d-1)\d^M_L\d^N_K.
\end{split}
\end{equation}
Here $d=11-D$ is the number of compact directions and $P_A{}^B{}_C{}^D$ is the projector on the adjoint representation of the corresponding duality group. It is defined as $P_A{}^B{}_C{}^DP_D{}^C{}_K{}^L=P_A{}^B{}_K{}^L$ and $P_A{}^B{}_B{}^A=\mbox{dim}(adj)$. The coefficients $\a_d$ and $\b_d$ depend on the duality group and for the cases in question take numerical values 
$(\a_4,\b_4)=(3,\fr{1}{3})$, $(\a_5,\b_5)=(4,\fr{1}{4})$, $(\a_6,\b_6)=(6,\fr{1}{3})$. The last line in \eqref{rel} with $D=\d^A_A$ is a direct consequence of the second relation and the properties of the projector. The first line is true only for $n\leq5$ and the relevant identity for $E_{6(6)}$ duality group reads
\beq
10P_Q{}^{(M}{}_T{}^NP_R{}^{P)}{}_S{}^{T}-P_R{}^{(M}{}_S{}^N\d^{P)}_Q-\fr13d^{MNP}d_{QRS}=0
\eeq

The invariant tensor plays very important role in the formalism of the extended geometry providing closure of the algebra of the generalised Lie derivatives. The closure constraint reads
\beq
\label{closure}
[\mc{L}_{V_1},\mc{L}_{V_2}]-\mc{L}_{[V_1,V_2]_C}=F_0,
\eeq
where the RHS is in general non-zero. The so-called \emph{section condition} assures that the algebra is closed i.e. $F_0=0$ and can be written as
\beq
Y_{KL}^{MN}\dt_M\bullet \dt_N\bullet =0,
\eeq
where $\bullet$ denotes any U-duality covariant expression. The Jacobiator of the transformations $\d_\S$ given by  \eqref{Lie} is zero up to the section condition as well.

The generalised metric is a dynamical field of the theory and along with its derivatives contributes to the effective potential \cite{Berman:2011jh, Hohm:2010pp}. This potential transforms under \eqref{Lie} as a scalar only if the section condition is satisfied. In the next sections we show that the Scherk-Schwarz reduction of the theory to five and six dimensions is consistent without the section condition constraint in the form in which it was formulated above.

\section{Scherk-Schwarz reduction}

In contrast to the Kaluza-Klein reduction here the dependence on internal coordinates is hidden in so-called twist matrices $W^{A}{}_{\bar{B}}(X)$ that are subject to various constraints. For the case at hand we consider the whole extended space as an internal space and let the remained $D$-dimensional space to be whatever it wants to be \cite{Aldazabal:2011nj, Grana:2012rr}:
\begin{equation}
\label{twist}
T^{A_1\ldots A_m}(X^M,x_{(D)})=W^{A_1}{}_{\bar{B}_1}(X)\cdots W^{A_m}{}_{\bar{B}_m}(X)T^{\bar{B}_1\ldots \bar{B}_m}(x_{(D)}),
\end{equation}
where $T$ is any generalised tensor with $m$ upper indices. Using inverse matrices $W_{bar{B}}{}^{A}$ similar expression can be written for a tensor with lower indices. From now on we will not include the dependence on $x_{(D)}$ since it doesn't affect the extended geometry formalism. The barred indices are the twisted ones (flat) and the unbarred are the untwisted ones (curved). To simplify notations we will use the unbarred indices for the flat space in cases where this does not confuse.

The important feature of the Scherk-Schwarz reduction is that it allows non-abelian gauge groups. Substituting the anzatz \eqref{twist} into the local transformations of the initial theory that are given by the generalised Lie derivative \eqref{Lie} we obtain the following transformation of the vector $Q^A$
\begin{equation}
\label{Lie_twisted}
\d_\S Q^{A}=(\mc{L}_\S Q)^A=W^{A}{}_{\bar{B}}X_{\br{K}{L}}{}^{\bar{B}}\S^{\bar{K}}Q^{\bar{L}}.
\end{equation}
Here the coefficients $X_{MN}{}^K$ are defined as
\begin{equation}
\label{Xgen}
X_{\br{A}{B}}{}^{\bar{C}}\equiv2W_C{}^{\bar{C}}\dt_{[\bar{A}}W_{\bar{B}]}{}^C+Y^{\br{C}{D}}_{\br{M}{B}}W_C{}^{\bar{M}}\dt_{\bar{D}}W_{\bar{A}}{}^{C}
\end{equation}
with the antisymmetrisation factor of $1/2$, and are assumed to be constants. One should note that in the case of extended geometry these ''structure constants'' are not antisymmetric.

We recall the closure constraint \eqref{closure}
\begin{equation}
\mc{L}_{[X_1,X_2]_C} Q^M -  [\mc{L}_{X_1},\mc{L}_{X_2}]Q^{M } = -  F_0^{M}.
\end{equation}
Assuming that $X_{MN}{}^K$ is constant and substituting the twist anzatz \eqref{twist} and the explicit form of $F_0$ \cite{Berman:2011jh, Berman:2011cg} this implies
\beq
\frac{1}{2} \left( X_{\bar{A} \bar{B} }{}^{\bar{C}} -    X_{\bar{B} \bar{A} }{}^{\bar{C} }\right)X_{\bar{C} \bar{E}}{}^{\bar{G}} - X_{\bar{B} \bar{E}}{}^{\bar{C}}X_{\bar{A} \bar{C}}{}^{\bar{G}} +   X_{\bar{A} \bar{E}}{}^{\bar{C}}X_{\bar{B} \bar{C}}{}^{\bar{G}}  = 0 \  
\eeq
for any $X_1$ and $X_2$. If we define $X_{MN}{}^K=(X_M)_N{}^K$ this can be written in the suggestive form
 \beq
  \label{closure1}
 [ X_{\bar{A}}, X_{\bar{B}}]  = - X_{[\bar{A} \bar{B}]}{}^{\bar{C}} X_{\bar{C}}  \ . 
 \eeq
This  allows one to interpret the structure constants as the components of the generators $X_M$ of the algebra of transformations
\beq
\label{G_transf}
\d_{\S}Q^{\bar{A}}=X_{\br{K}{L}}{}^{\bar{A}}\S^{\bar{K}}Q^{\bar{L}}
\eeq
in adjoint representation. By making use of the closure constraint \eqref{closure1} we find the Jacobiator
\beq
\begin{split}
\label{Jac}
&[\delta_{\S_1}, [\delta_{\S_2}, \delta_{\S_3} ] ]V^{\bar{F}} + c.p. = \\ & \left( X_{[\bar{A}\bar{B}]}{}^{\bar{E}}X_{[\bar{E} \bar{C}] }{}^{\bar{G}}   +   X_{[\bar{C}\bar{A}]}{}^{\bar{E}}X_{[\bar{E} \bar{B}] }{}^{\bar{G}} +X_{[\bar{B}\bar{C}]}{}^{\bar{E}}X_{[\bar{E} \bar{A}] }{}^{\bar{G}}   \right)X_{\bar{G} \bar{D}}{}^{\bar{F} } \S_1^{\bar{A}}\S_2^{\bar{B}} \S_3^{\bar{C}} V^{\bar{D}},
\end{split}
\eeq
where {\it c.p.} denotes cyclic permutations. The right hand side of this equation is the Jacobi identity of the antisymmetric part $X_{[MN]}{}^K$ projected into the algebra generator. For the consistency of the algebra of transformations the right hand side should vanish. We emphasise that the Jacobi identity for $X_{[MN]}{}^K$ needs only to hold after the projection.

We need $X_{MN}{}^K$ to be not only constants but also invariant objects under the local symmetry transformations. As it will be shown later it is necessary so that the reduced effective potential does not depend on the internal coordinates and transforms as a scalar. As it follows from the definition \eqref{Lie_twisted} the structure constants $X_{MN}{}^K$ should transform as a generalised tensor
\beq
\delta_{\S} X_{\bar{A} \bar{B}}{}^{\bar{C}}  =   \S^{\bar{E}} \left(  [ X_{\bar{E}}, X_{\bar{A}}] _{\bar{B}}{}^{\bar{C}} +  X_{\bar{E} \bar{A}}{}^{\bar{D}} (X_{\bar{D}})_{\bar{B}}{}^{\bar{C}} \right).
\eeq
This leads to the final quadratic constraint on the structure constants
 \beq
  \label{closure2}
 [ X_{\bar{A}}, X_{\bar{B}}]  = - X_{\bar{A} \bar{B}}{}^{\bar{C}} X_{\bar{C}}  \ . 
 \eeq
We conclude from this constraint that the symmetric part $Z_{MN}{}^K=X_{(MN)}{}^K$ should vanish when projected into a generator
\beq
Z_{\bar{A} \bar{B}}{}^{\bar{C}} X_{\bar{C}}  = 0 \ .
\eeq

The quadratic constraint \eqref{closure2} on its own is enough to ensure that the Jacobiator \eqref{Jac} vanishes and the algebra is closed. This can be seen by considering the Jacobi identity for the commutator appearing in \eqref{closure2}. Hence the closure condition can be relaxed from the section condition that restricts fields and their products to a condition on the structure constants $X_{MN}{}^K$ that define the algebra of gauge transformations.

\section{The embedding tensor formalism in gauged supergravity} 
 Toroidal compactification of eleven-dimensional supergravity gives rise to maximal supergravities in $D=11- d$ dimensions which admit a global $G=E_{d(d)}$ duality (or Cremmer-Julia) symmetry.  These theories allow SUSY preserving deformations, known as gaugings, in which some subgroup of the global $E_{d(d)}$ symmetry is promoted to a local symmetry.  The resultant gauged supergravities have non-abelian gauge groups and develop a potential for the scalar fields.    A  universal approach to gauged supergravities is the embedding tensor (for a review see \cite{Samtleben:2008pe}) which describes how the gauge group generators are embedded into the global symmetry.  Treated as a spurionic object the embedding tensor provides a manifestly $G$ covariant description of the gauged supergravities. 

  In addition to the global $E_{d(d)}$ symmetry the toroidally reduced theories also posses a global $\mathbb{R}^+$ scaling symmetry  known as the trombone symmetry (this is an on-shell symmetry for $D\neq2$).   This gives rise to a more general class of gaugings whereby a subgroup of the full global duality group $E_{d(d)}\times \mathbb{R}^+$  is promoted to a local symmetry.  The embedding tensor approach was extended to incorporate such trombone gaugings in \cite{LeDiffon:2008sh}.
  
 In this paper we consider $D=6$ \cite{Bergshoeff:2007ef} and $D=5$ \cite{deWit:2004nw} cases for which the vector fields of the un-gauged/abelian theory are in the spinor representation $\bf 16$ of $SO(5,5)\equiv E_{5(5)}$ and the fundamental representation $\bf 27$ of $E_{6(6)}$ respectively. For further reference we denote these representations as $\mc{R}_V$. The gaugings are specified by the embedding tensor $\widehat{\Theta}_M{}^{\underline{\a}}$ that  projects generators  $t_{\underline{\a}}$ of the global duality group $E_{d(d)}\otimes\mathbb{R}^+$ to some subset $X_{M}=\widehat{\Q}_{M}{}^{\underline{\a}} t_{\underline{\a}}$ which generate the gauge group and enter into covariant derivatives:
  \beq
  D = \nabla - g A^{M} X_{M} \ .
    \eeq 
The index $\underline{\a}$ of the embedding tensot is a multiindex which labels the adjoint representation of the duality group. According to its index structure the embedding tensor is in the $\mc{R}_V\times \mc{R}_{adj}$ representation, where $\mc{R}_{adj}$ is the adjoint representation of the global duality group. In general it decomposes as
 \beq
 \widehat{\Q}_M{}^{\underline{\a}} \in \mc{R}_V\otimes\mc{R}_{adj}=\mc{R}_V\oplus\ldots
 \eeq
The preservation of supersymmetry gives a linear constraint restricting the embedding tensor only to 
\beq
\begin{split}
&\widehat{\Q}_M{}^{\underline{\a}} \in {\bf 16}_s\oplus {\bf 144}_c\mbox{, for $D=6$} \\
&\widehat{\Q}_M{}^{\underline{\a}} \in \bf{27\oplus 351}\mbox{, for $D=5$}.
\end{split}
\eeq
The trombone gauging that is always in the representation $\mc{R}_V$ corresponds to the on-shell symmetry and doesn't appear in the potential. The scalar potentials written further below do not include the trombone. In the next two subsection we briefly review the structure of the scalar sector of the maximal gauged supergravities in $D=5,6$ and introduce expressions that we need for the further sections. Since the review is very brief and doesn't cover all the details we refer the reader to the relevant papers \cite{deWit:2004nw, Bergshoeff:2007ef, LeDiffon:2008sh} and \cite{Samtleben:2008pe}.

\subsection{$D=6$ supergravity}     

Maximal supergravity in six dimensions is invariant under the global duality group $SO(5,5)$. The representation $\mc{R}_V$ is now the spinorial representation ${\bf 16}_s$ of $SO(5,5)$. We let the capital Latin indices run from 1 to 16 labelling this representation and the small Latin indices run from 1 to 10 labelling the $\bf{10}$ representation of $SO(5,5)$. Then the components of the projected generators $X_M$ can be written in the spinorial representation as
\beq
X_{MN}{}^K=(X_M)_N{}^K=\widehat{\Q}_M{}^{ij} (t_{ij})_N{}^K=\widehat{\Q}_M{}^{ij}(\G_{ij})_N{}^K,
\eeq
where $\G_{ij}=\G_{[i}\G_{j]}$ are the generators $t_{ij}$ in the spinorial representation while $\G_i$ are $16\times16$ gamma matrices in the Majorana representation. This means that they are real and symmetric
\beq
\G_i{}^{MN}=\G_i{}^{NM}.
\eeq
As it was shown in \cite{Bergshoeff:2007ef} and \cite{LeDiffon:2008sh} the gauge group generators are given by 
\beq
\label{X6}
X_{MN}{}^K=-\q^{iL}\G^j{}_{LM}(\G_{ij})_{N}{}^K-\fr{1}{10}(\G^{ij})_M{}^L(\G_{ij})_N{}^K\q_L-\q_M\d_N{}^K.
\eeq
The generators are only written in terms of the gauging $\q^{iM}\in{\bf 144}$ and the trombone gauging $\q_M\in{\bf 16}$. The symmetric part $Z_{MN}{}^K=X_{(MN)}{}^K$ then reads
\beq
Z_{MN}{}^K=\G_{iMN}\hat{Z}^{iM},\quad Z^{iM}=-\q^{iM}-\fr25\G^{iMN}\q_N.
\eeq
Since the gauging $\q^{iM}$ is in the $\bf 144$ representation it satisfies the linear constraint \\ $\q^{iM}\G_{iMN}~=~0$.

Scalar fields of the theory are elements of the coset space $SO(5,5)/SO(5)\times SO(5)$ that can be conveniently parametrised by $SO(5,5)$ valued $16\times 16$ matrices $V_M{}^{\a\dot{\a}}$ \cite{Tanii:1984zk}. its inverse is defined by
\beq
V_{M}{}^{\a\dot{\a}}V^N{}_{\a\dot{\a}}=\d_M{}^N,\quad V_{M}{}^{\a\dot{\a}}V^M{}_{\b\dot{\b}}=\d^\a_\b\d^{\dot{\a}}_{\dot{\b}}.
\eeq
Here the dotted and the undotted small Greek indices run from 1 to 4 and label the spinor representation $\bf 4$ of each $SO(5)$ in the coset. 

In the absence of the trombone gauging the scalar potential can be written as
\beq
V_{scalar}=g^2\Tr\left(T^{\hat{a}}\widetilde{T}^{\hat{a}}-\fr12T\widetilde{T}\right),
\eeq
where tilde denotes transposition and the $T$-tensors are given by \cite{Bergshoeff:2007ef}
\beq
\label{T_tensor}
\begin{split}
(T^{\hat{a}})^{\a\dot{\a}}&=\mc{V}_i{}^{\hat{a}}\q^{iM}V_M{}^{\a\dot{\a}}\\
(T^{\hat{\dot{a}}})^{\a\dot{\a}}&=-\mc{V}_i{}^{\hat{\dot{a}}}\q^{iM}V_M{}^{\a\dot{\a}}\\
T&=T^{\hat{a}}\g^{\hat{a}}=-T^{\hat{\dot{a}}}\g^{\hat{\dot{a}}}.
\end{split}
\eeq
Here the hatted small Latin indices label the vector representation $\bf{5}$ of $SO(5)$ and dots again distinguish between two $SO(5)$'s in the coset. The gamma matrices $\gamma^{\hat{a}}$  and $\gamma^{\hat{\dot{\a}}}$ are $4\times4$ chiral gamma matrices whose vector indices are contracted without raising and lowering. The $10\times 5$ matrices $\mc{V}$ are defined as
\beq
\begin{split}
&\mc{V}_i{}^{\hat{a}}=\fr{1}{16}V_M{}^{\a\dot{\a}}(\g^{\hat{a}})_{\a}{}^{\b}\G_i{}^{MN}V_N{}_{\b\dot{\a}},\\ &\mc{V}_i{}^{\hat{\dot{a}}}=-\fr{1}{16}V_M{}^{\a\dot{\a}}(\g^{\hat{\dot{a}}})_{\dot{\a}}{}^{\dot{\b}}\G_i{}^{MN}V_N{}_{\a\dot{\b}}.
\end{split}
\eeq
According to the quadratic constraint the dotted and the undotted $T$ tensors are not independent and satisfy
\label{TT}
\beq
T^{\hat{a}}{}_{\a\dot{\a}}T^{\hat{a}}{}_{\b\dot{\b}}=T^{\hat{\dot{a}}}{}_{\a\dot{\a}}T^{\hat{\dot{a}}}{}_{\b\dot{\b}}.
\eeq

\subsection{$D=5$ supergravity}     
     
In five dimensions the global duality group of the maximal supergravity is $E_{6(6)}$ that is the maximal real subgroup of the complexified $E_6$ group. The representation $\mc{R}_V$ in this case is given by the $\bf 27$ representation of $E_{6(6)}$ and the capital Latin indices run from 1 to 27. The corresponding invariant tensor is a fully symmetric tensor $d_{MNK}$ that satisfies the following identities 
\begin{equation}
\label{d_ident0}
\begin{split}
d_{MPQ}d^{NPQ}&=\d_M^N,\\
d_{MRS}d^{SPT}d_{TNU}d^{URQ}&=\fr{1}{10}\d_{(M}^{P}\d_{N)}^{Q}-\fr25d_{MNR}d^{RQP},\\
d_{MPS}d^{SQT}d_{TRU}d^{UPV}d_{VQW}d^{WRN}&=-\fr{3}{10}\d_M^N.
\end{split}
\end{equation}

The linear constraint implied by supersymmetry restricts the full embedding tensor $\widehat{\Q}_M{}^{\underline{\a}}$ to the $\bf 27\oplus 351$ representation of $E_{6(6)}$. In the absence of the trombone gauging the embedding tensor reads
\beq
\Q_M{}^{\underline{\a}}=Z^{PQ}(t^{\underline{\a}})_R{}^Sd^{RKL}d_{MNK}d_{SQL}.
\eeq
The symmetric part of the gauge group generators $Z_{MN}{}^K=X_{(MN)}{}^K$ is then given by 
\beq
Z_{MN}{}^K=d_{MNL}\hat{Z}^{KL}, \quad \hat{Z}^{KL}=Z^{KL}-\fr{15}{2}d^{KLM}\q_{M}.
\eeq
A non-trivial relation among the generators of $E_{6(6)}$ that follows from the last line in \eqref{rel} is
\beq
(t^{\underline{\a}})_M{}^K(t_{\underline{\a}})_N{}^L=\fr{1}{18}\d_M^K\d_N^L+\fr16\d_M^L\d_N^K-\fr53d_{MNR}d^{RKL}.
\eeq

Scalar fields of the theory live in the coset space $E_{6(6)}/USp(8)$ and can be pa\-ra\-met\-ri\-sed by the scalar matrix $\mc{V}_M{}^{ij}$ with small Latin indices labelling the $\bf 8$ representation of $USp(8)$. The scalar matrix $\mc{V}_M{}^{ij}$ is antisymmetric in $ij$ and satisfies $\mc{V}_M{}^{ij}\W_{ij}=0$, where $\W_{ij}=-\W_{ji}$ is the symplectic invariant of $USp(8)$. Thus, the scalar matrix has $27\times 27$ components and its inverse is defined as
\beq
\begin{split}
&\mc{V}_M{}^{ij}\mc{V}_{ij}{}^N=\d_M^N\\
&\mc{V}_{ij}{}^M\mc{V}_M{}^{kl}=\d_{ij}{}^{kl}-\fr18\W_{ij}\W^{kl}.
\end{split}
\eeq

The matrix $\mc{V}$ can be used to elevate the embedding tensor to the so-called $T$-tensor that is $USp(8)$ covariant field dependent tensor. We need this tensor since it appears in the scalar potential. The convenient relation to be exploited  below is \cite{deWit:2004nw}
\beq
\label{XT}
X_{MN}{}^P=\mc{V}_{M}{}^{mn}\mc{V}_N{}^{kl}\mc{V}_{ij}{}^P\left[2\d_k{}^iT^j{}_{lmn} + T^{ijpq}{}_{mn}\W_{pk}\W_{ql}\lefteqn{\ph{\fr12}}\right]
\eeq
The tensor $T^{klmn}{}_{ij}$ belongs to the $\bf 315$ representation while $T^i{}_{jlm}$ is in the $\bf 36\oplus 315$. It is possible to write these two tensors in terms of two pseudoreal, symplectic traceless, tensors $A_1{}^{ij}\in\bf 35$ and $A_2{}^{i,jkl}\in \bf 315$ as
\beq
\label{TA}
\begin{split}
T^{klmn}{}_{ij}&=4A_2{}^{q,[klm}\d^{n]}{}_{[i}\W_{j]q}+3A_2{}^{p,q[kl}\W^{mn]}\W_{p[i}\W_{j]q},\\
T_i{}^{jkl}&= -\W_{im}A_2{}^{(m,j)kl}-\W_{im}\left(\W^{m[k}A_1{}^{l]j}+\W^{j[k}A_1{}^{l]j}+\fr14\W^{kl}A_1{}^{mj}\right).
\end{split}
\eeq
Tensors $A_1$  and $A_2$ satisfy $A_1{}^{[ij]}=0$, $A_2{}^{i,jkl}=A_2{}^{i,[jkl]}$ and $A_2{}^{[i,jkl]}=0$. The scalar potential then can be written as
\beq
V_{scalar}=g^2\left[3|A_1{}^{ij}|^2-\fr13|A_2{}^{i,jkl}|^2\right],
\eeq
where $|\;|^2$ stands for the contraction of all indices.

\section{Algebraic structure}

The general form of the structure constants $X_{AB}{}^{C}$ is always the same and is given by \eqref{Xgen}. Under a particular U-duality group these split into certain representations that depend on the duality group and are identified with gaugings. In this section we give an explicit derivation of the embedding tensor and all gaugings starting from $X_{AB}{}^C$ in its general form.

\subsection{$SO(5,5)$: reduction to 6 dimensions }

Maximal supergravity in 6 dimensions possesses a global duality group $E_{5(5)}=SO(5,5)$. The local group of the theory is $SO(5)\times SO(5)$. Thus the target space of scalar fields of the theory is given by the coset
\begin{equation}
\frac{SO(5,5)}{SO(5)\times SO(5)}.
\end{equation}
The corresponding extended space of the Berman-Perry formalism has 16 dimensions and the representation $\mc{R}_V$ appears to be the spinorial representation of $SO(5,5)$.

The invariant tensor of the duality group is given by the contraction of two gamma matrices in the Majorana representation 
\begin{equation}
Y^{MN}_{KL}=\fr{1}{2}\G^{iMN}\G_{iKL},
\end{equation}
that are thus symmetric and real. Here capital Latin indices run from 1 to 16 and small Latin indices run from 1 to 10 labelling the vector representation of $SO(5,5)$. Since the generators $t^{\underline{\a}}$ of the duality group  in the spinorial representation are given by $\G^{ij}$, where the multiindex $\underline{\a}$ is represented by the antisymmetric pair of vector indices, the projector with correct normalisation is defined as
\begin{equation} 
P_N{}^M{}_L{}^K=-\fr{1}{32}(\G^{ij})_N{}^M(\G_{ij})_L{}^K.
\end{equation}

All gaugings of the maximal supergravity appear as components of the structure constants (or the embedding tensor). Start with the trace part of the structure constants \eqref{Xgen}
\begin{equation}
\label{XtraceSO}
X_{\br{M}{N}}{}^{\bar{N}}=4\dt_{C}W^C_{\bar{M}} +W^{\bar{C}}_C\dt_{\bar{M}}W^C_{\bar{C}} =:-16\q_{\bar{M}}.
\end{equation}
By making use of the algebra of gamma matrices the symmetric part of the gaugings can be extracted as
\begin{equation}
\label{ZSO}
\begin{split}
X_{(AB)}{}^C&=\G_{iAB}Z^{iC},\\
Z^{\br{i}{C}}&=\fr14 \G^{\bar{j}\br{C}{D}}G_{\bar{j}}{}^i\dt_{\bar{D}}G_{i}{}^{\bar{i}},
\end{split}
\end{equation}
where the twist matrices in the vector representation $G_{i}{}^{\bar{j}}$ are defined as
\begin{eqnarray}
\G^{iAB}G_{i}{}^{\bar{j}}=\G^{\bar{j}\br{C}{D}}W_{\bar{C}}{}^AW_{\bar{D}}{}^B.
\end{eqnarray}
According to its indices the gauging $Z^{iB}$ is in the $\bf{16\otimes10=16\oplus144}$ representation of $SO(5,5)$. Separating the $\bf{16}$ part of the gauging we obtain the trombone gauging $\q_M$
\begin{equation}
Z^{iM}\G_{iMN}=-4\q_N.
\end{equation}
What is left lives in the $\bf{144}$ representation and is defined as
\begin{equation}
\q^{iM}=-Z^{iM}-\fr25\G^{iMN}\q_N.
\end{equation}
After some algebra (see Appendix A) the structure constants can be rewritten in terms of only these objects 
\begin{equation}
\label{XSO}
X_{MN}{}^K= -\q^{iL}\G^j_{LM}(\G_{ij})_N{}^K - \fr{1}{10}(\G^{ij})_M{}^L(\G_{ij})_N{}^K\q_L - \d_N^K\q_M.
\end{equation}
This has exactly the same structure as the embedding tensor of the maximal supergravity in 6 dimensions \eqref{X6}
\begin{equation}
X_{MN}{}^K \in \bf{16\oplus 144}.
\end{equation}
with gaugings explicitly written in terms of the twist matrices as
\beq
\label{gaugingsSO}
\begin{split}
\q^{\br{i}{M}}&=-\fr14 \G^{\bar{j}\br{M}{D}}G_{\bar{j}}{}^i\dt_{\bar{D}}G_{i}{}^{\bar{i}}-\fr25\G^{\br{i}{M}\bar{N}}\q_{N},\\
\q_{\bar{N}}&=-\fr{1}{16}\G^{\br{A}{D}\bar{i}}\G_{\br{j}{A}\bar{N}}G_{\bar{i}}{}^{i}\dt_{\bar{D}}G_{i}{}^{\bar{i}}.
\end{split}
\eeq
It is straightforward to check that the second line here is the same as \eqref{XtraceSO} using the definition of $G_{i}{}^{\bar{i}}$ and the relation
\beq
G_{\bar{j}}{}^j\dt_{\bar{A}}G_j{}^{\bar{i}}=\fr18\G^{\br{B}{K}\bar{i}}\G_{\bar{j}\br{C}{K}}W_C{}^{\bar{C}}\dt_{\bar{A}}W_{\bar{B}}{}^C.
\eeq

\subsection{$E_{6(6)}$: reduction to 5 dimensions }

In five dimensions vector fields of maximal supergravity transform in the $\mathbf 27$ representation of the global duality group $E_{6(6)}$. The scalar fields transform non-linearly and are parametrised by elements of the coset
\begin{equation}
\frac{E_{6(6)}}{USp(8)}.
\end{equation}
The group $USp(8)$ is the R-symmetry group of the theory. 

The U-duality invariant formalism of the extended geometry provides the extended space to be 27 dimensional and the generalised vector indices $A, B \ldots$ label the $\mathbf{27}$ representation of $E_{6(6)}$. The invariant tensor is given by the $E_{6(6)}$ symmetric invariant tensor $d_{MNK}$
\begin{equation}
Y^{MN}_{RS}=10d^{MNK}d_{KRS},
\end{equation}
that is subject to the following useful identities
\begin{equation}
\label{d_indent}
\begin{split}
d_{MPQ}d^{NPQ}&=\d_M^N,\\
d_{MRS}d^{SPT}d_{TNU}d^{URQ}&=\fr{1}{10}\d_{(M}^{P}\d_{N)}^{Q}-\fr25d_{MNR}d^{RQP},\\
d_{MPS}d^{SQT}d_{TRU}d^{UPV}d_{VQW}d^{WRN}&=-\fr{3}{10}\d_M^N.
\end{split}
\end{equation}
The trace part of the structure constant \eqref{Xgen} is identified with the trombone gauging $\q_{M}$ and reads
\begin{equation}
\label{XtraceE}
X_{\br{M}{N}}{}^{\bar{N}}=9\dt_{C}W^C_{\bar{M}}+W^{\bar{C}}_C\dt_{\bar{M}}W^C_{\bar{C}}=-27\q_{\bar{M}}.
\end{equation}
The intertwining tensor is given by the symmetric part of the structure constant $Z_{MN}{}^K=X_{(MN)}{}^K$ and is parametrised by the tensor $\hat{Z}^{MN}$ in the $\mathbf{27\oplus351}$
\begin{equation}
\begin{split}
&Z_{MN}{}^K=d_{MNR}\hat{Z}^{RK}
\end{split}
\end{equation}
Taking the symmetric part of \eqref{Xgen} and by making use of the identities \eqref{d_indent} we have for the symmetric part
\begin{equation}
\label{ZE}
\hat{Z}^{\br{M}{N}}=5d^{\br{M}{K}\bar{L}}W^{C}_{\bar{L}}\dt_{\bar{K}}W^{\bar{N}}_C
\end{equation}
that has the same structure as \eqref{ZSO} if one notices that
\begin{eqnarray}
W^{M}_{\bar{M}}d^{\bar{M}\br{K}{L}}=d^{MKL}W_{K}^{\bar{K}}W_{L}^{\bar{L}}
\end{eqnarray}
since the twist matrices interpolate between the barred and the unbarred indices. 

Subtracting the part of the tensor \eqref{ZE} that is symmetric in $MN$ we are left with the gauging in the $\mathbf{351}$ and the trombone gauging
\begin{equation}
\begin{split}
Z^{MN}&=\hat{Z}^{MN}+\fr{15}{2}d^{MNK}\q_{K},\\
\q_{\bar{N}}&=5\,d^{\br{M}{B}\bar{L}}d_{\br{M}{N}\bar{K}}W^L_{\bar{L}}\dt_{\bar{B}}W^{\bar{N}}_L.
\end{split}
\end{equation}
Thus the structure constant $X_{MN}{}^K$ is in the $\mathbf{27\oplus351}$ representation of $E_{6(6)}$.

%\subsection{$SL(5)$ revisited}

\section{Scalar potential}

The effective potential $V=V(M_{AB}, \dt_KM_{AB})$ that depends on the generalised metric $M_{AB}$ and its derivatives after twisting should become the scalar potential for the appropriate gauged SUGRA. It appears that one must add an extra term of type
\begin{equation}
Y^{AB}_{MN}\dt_{A}E_{\X}{}^M\dt_BE_{\Q}{}^N\d^{\X\Q},
\end{equation}
where $E_\X{}^M$ is a generalised vielbein and $\d^{\X\Q}$ is the Kronecker delta. It can be always added to the scalar potential since it is zero up to the section condition. This term is necessary for two major reasons. Firstly, this term allows to write the potential in terms of the structure constants $X_{MN}{}^K$. Secondly, this terms provides the potential that is invariant under gauge transformations \eqref{G_transf}. Not to be confused, one should think of this term as a term that \emph{has always been in the potential} but has usually been dropped because of the section condition. After the section condition is relaxed it is important to add this term since it provides the invariance of the potential.

\subsection{$D=6$ supergravity}

The effective potential in six dimensions is given by \cite{Berman:2011jh}
\begin{equation}
\label{VSO}
\begin{split}
V_{eff}=&\fr{1}{16}M^{MN}\dt_M M^{KL}\dt_N M_{KL} -\fr12 M^{MN}\dt_N M^{KL} \dt_L M_{NK} +\\
  &+\fr{11}{1728} M^{MN} (M^{KL}\dt_MM_{KL})(M^{RS}\dt_NM_{RS}) +2 Y^{MN}_{KL}\dt_M E_\Q{}^K\dt_NE_\X{}^L\d^{\X\Q}.
\end{split}
\end{equation}
Here the $16\times16$ matrix $M_{KL}$ is the generalised metric and it is written in terms of the metric $g_{\m\n}$ and the RR 3-form field $C_{\m\a\b}$
\begin{equation}
M=\fr{1}{\sqrt{g}}
\begin{pmatrix} g_{\mu \nu}+{\fr12} C_{\mu}{}^{\r\s} C_{\nu \r\s} +
{\fr{1}{16}}    X_{\mu}  X_{\nu}  &
{\fr{1}{\sqrt{2}}} C_{\mu}{}^{ \nu_1 \nu_2} + {\fr{1}{4 \sqrt
{2}}} X_{\mu} V^{\nu_1 \nu_2} & {\fr{1}{4}} {g}^{-{1 / 2}} X_{\mu} \\
{\fr{1}{\sqrt{2}}} C^{\mu_1 \mu_2}{}_{\nu} + {\fr{1}{4 \sqrt
{2}}}  V^{\mu_1 \mu_2} X_{\nu}  & g^{\mu_1 \mu_2,\nu_1 \nu_2}+ {\fr{1}
{2}} V^{\mu_1 \mu_2}V^{\nu_1 \nu_2} & {\fr{1}{\sqrt{2}}}  g^{-
{1 / 2}} V^{\mu_1 \mu_2} \\
{\fr{1}{4}} g^{-{1 / 2}} X_{\nu} & {\fr{1}{\sqrt{2}}} g^{-{1 /
2}} V^{\nu_1 \nu_2} & g^{-1}
\end{pmatrix},
\end{equation}
where the small Greek letters here run from 1 to 5 labelling 5 compact directions and
\beq
V^{\r\s}={\fr{1}{3!}} \epsilon^{\r\s\m\n\w}C_{\m\n\w}, \qquad \qquad X_{\m}=C_{\m\n\r} V^{\n\r}.
\eeq

The matrix $E_\X{}^K$ is the vielbein for $M^{MN}=E_\Q{}^ME_\X{}^N\d^{\Q\X}$ and the capital Greek indices run from 1 to 16 labelling flat spinorial indices.

For the convenience of notations we define the object
\begin{equation}
\label{f}
f_{\br{A}{B}}{}^{\bar{C}}=W_C{}^{\bar{C}}\dt_{\bar{A}}W^{C}{}_{\bar{B}},
\end{equation}
where $W^{C}{}_{\bar{B}}$ is the twist matrix introduced in \eqref{twist}. Then using the definition \eqref{Xgen} the structure constant can be written as
\begin{equation}
\label{X2f}
X_{MN}{}^K=f_{MN}{}^K-f_{NM}{}^K+Y^{KL}_{BN}f_{LM}{}^B,
\end{equation}
that us true for the extended geometry formalism in any dimension. 

From now on we assume that the trombone gauging vanishes and that the matrix $M_{MN}$ is unimodular. The latter can be always arranged by rescaling the generalised metric by $g=\mbox{det}(g_{\m\n})$. The only effect this has on the potential is change in the coefficients of the terms proportional to derivatives of the determinant. Summarising we have
\begin{equation}
\begin{split}
\q_M&=0, \quad\mbox{det} W=1,\\
f_{AB}{}^A&=0, \quad f_{A B}{}^B=0,\\
\dt_C W^C{}_{\bar{B}}&=0.
\end{split}
\end{equation}
In cases when it doesn't confuse the reader the bar notation is dropped to make expression less heavy. In all expressions which include terms with both barred and unbarred indices these are treated carefully. One should remember that such quantities like $X_{MN}{}^K$, $f_{MN}{}^K$ or gaugings always have flat barred indices and not be confused if they appear without bar. Taking this into account, the effective potential is given by
\begin{equation}
V_{eff}=V_1 + V_2 + V_3 + SC,
\end{equation}
where
\begin{equation}
\begin{split}
V_1 &= -\fr18M^{MN}f_{NP}{}^Lf_{ML}{}^P+M^{MN}f_{MP}{}^Lf_{LN}{}^P,\\
V_2 &= \fr12M^{MN}f_{PM}{}^Lf_{LN}{}^P,\\
V_3 &= M^{MN}M^{KL}M_{RS}\left(\fr18 f_{MK}{}^Rf_{NL}{}^S-\fr12f_{KM}{}^Rf_{NL}{}^S\right),\\
SC  &= \fr12 Y^{MN}_{KL}f_{MR}{}^Kf_{NS}{}^L M^{RS}.
\end{split}
\end{equation}
By integrating $\dt_P$ and $\dt_L$ by part in $V_2$ it can be shown that this term is zero up to a full derivative.
To proceed further and to be able to use gamma matrices algebra we need to define objects in the vector representation
\begin{equation}
\label{Gm}
\begin{split}
f_A{}_j{}^i&=\fr18(\G_j{}^i)_K{}^Lf_{AK}{}^L,\\
m_{\br{i}{j}}\G^{\bar{j}\br{A}{B}}&=\G_{\bar{i}\br{R}{S}}M^{\br{R}{A}}M^{\br{S}{B}},\\
X_M{}_i{}^j&=\fr18(\G_i{}^j)_K{}^LX_{ML}{}^K
\end{split}
\end{equation}
By making use of these definitions the part $V_3$ can be written as
\begin{equation}
\label{V3}
\begin{split}
V_3=&\fr14(\G_b\G^n)^N{}_Lf_K{}_i{}^jf_N{}_m{}^bm^{im}m_{jn}M^{KL}=\\
&\fr{1}{16}X_M{}_i{}^jX_N{}_k{}^lM^{MN}m^{ik}m_{jl}=\fr{1}{32} X_{MR}{}^KX_{NS}{}^LM^{MN}M^{RS}M_{KL}.
\end{split}
\end{equation}
Indeed, the first two lines of \eqref{Gm} imply that
\begin{equation}
\begin{aligned}
&f_{MK}{}^Rf_{NL}{}^SM_{RS}M^{MN}M^{KL}=2f_M{}_i{}^jf_N{}_m{}^nM^{MN}m_{nj}m^{mi},\\
&f_{KM}{}^Rf_{NL}{}^SM_{RS}M^{MN}M^{KL}=\fr12(\G^n{}_b)^N{}_Lf_K{}_i{}^jf_N{}_m{}^bm^{im}m_{jn}M^{KL}.
\end{aligned}
\end{equation}
These two equalities lead to the first line in \eqref{V3}. The definitions \eqref{f} and \eqref{Xgen} together with the condition $\q_M=0$ allow to write the structure constants $X_{MN}{}^K$ in terms of $f_{AB}{}^L$
\begin{equation}
\begin{aligned}
4Z^{iC}&=&&\G^{iAB}f_{AB}{}^C,\\
X_{MN}{}^K&=&&\fr14\G^{iAB}\G^j{}_{LM}(\G_{ij})_N{}^Kf_{AB}{}^L.
\end{aligned}
\end{equation}
Note, that this relation can not be inverted i.e. it is impossible to write $f_{AB}{}^C$ in terms of $X_{MN}{}^K$ and just substitute it into the potential. Basically, this follows from the first line of the equation above, that 
includes only symmetric part. Finally, substituting the last line of the equation \eqref{Gm} into the second line of \eqref{V3} and using the identities above one exactly recovers the first line in \eqref{V3}.

To obtain the term $V_1+SC$ one may use the following relations
\beq
\label{YXSO}
\begin{aligned}
Y_{SM}^{RL}X_{KR}{}^S&=-3X_{KM}{}^L,\\
Y_{SM}^{RL}f_{KR}{}^S&=-3f_{KM}{}^L,
\end{aligned}
\eeq
that follow from the explicit form of the structure constant \eqref{XSO}, relation between $X_{MN}{}^K$ and $f_{MN}{}^K$ \eqref{X2f}, identities \eqref{rel} involving the invariant tensor $Y_{MN}^{KL}$ and the condition $\q_M=0$. Then the term $V_1+SC$ of the effective potential can be written as
\beq
V_1+SC=-\fr{1}{8}X_{MK}{}^LX_{NL}{}^KM^{MN}.
\eeq
Indeed, substituting \eqref{X2f} into the expression above one encounters exactly $V_1+SC$ plus a term, proportional to $V_2$, that is a full derivative. 

Finally, the effective potential can be recast in the following form
\begin{equation}
\label{VSOt}
V_{eff}=-\fr{1}{8}X_{MK}{}^LX_{NL}{}^KM^{MN}+\fr{1}{32}X_{MR}{}^KX_{NS}{}^LM^{MN}M^{RS}M_{KL}.
\end{equation}
This expression reproduces exactly the scalar potential for maximal gauged supergravity in $D=6$ dimensions up to a prefactor
\begin{equation}
V_{eff}=6\Tr\left[T^{\hat{a}}\tilde{T}^{\hat{a}}-\fr12T\tilde{T}\right]=6V_{scalar}.
\end{equation}
The details of this calculation are provided in Appendix B.1.

The effective potential \eqref{VSOt} is invariant under transformations \eqref{G_transf} because of the quadratic constraint \eqref{closure2} (see Appendix B).

\subsection{$D=5$ supergravity}

The low energy effective potential for the $E_{6(6)}$ invariant M-theory has the same form as in the $SO(5,5)$ case up to coefficients \cite{Berman:2011jh}
\begin{equation}
\label{VE}
\begin{split}
V_{eff}=&\fr{1}{24}M^{MN}\dt_M M^{KL}\dt_N M_{KL} -\fr12 M^{MN}\dt_N M^{KL} \dt_L M_{NK} +\\
  &+\fr{19}{9720} M^{MN} (M^{KL}\dt_MM_{KL})(M^{RS}\dt_NM_{RS}) -\fr12Y^{MN}_{KL}\dt_M E_\Q{}^K\dt_NE_\X{}^L\d^{\X\Q}.
\end{split}
\end{equation}
We again add the term proportional to the section condition that includes the vielbein
\begin{equation}
E_{\Q}{}^{M} =  (\textup{det}e)^{-1/2}
\begin{pmatrix}
e_{\mu}{}^{i} & - {\fr{1}{\sqrt{2}}} e_{\mu}{}^{j} C_{j i_1 i_2} &
{\fr{1}{2}} e_{\mu}{}^{i_3} U + {\fr{1}{4}} e_{\nu}{}^{i_3} C_
{\mu j k} V^{\nu j k} \\
0 & e^{\mu_1}{}_{[i_1} e^{\mu_{2}}{}_{i_2]} & -{\fr{1}{\sqrt{2}}} e^
{\mu_1}{}_{j_1} e^{\mu_{2}}{}_{j_2} V^{j_1 j_2 i_3} \\
0 & 0 & (\textup{det}e)^{-1} e_{\mu_3}^{i_3}
\end{pmatrix},
\end{equation}
where the capital Greek letters now run from 1 to 27, the small Latin and Greek indices run from 1 to 6 labelling  curved and flat space respectively. The fields $U$ and $V^{ijk}$ are defined as
\beq
\begin{split}
U&=\fr{1}{6}\e^{ijklmn}C_{ijklmn},\\
V^{ikl}&=\fr{1}{3!}\e^{iklmnj}C_{mnj}.
\end{split}
\eeq
Here the 6-form field $C_{ijklmn}$ is a new field that was not present in the previous example because the dimension was lower than 6.

Using the same notations for $f_{MN}{}^K$ as in the previous subsection and setting $\det M=1$ and $\q_M=0$ we have for the twisted effective potential
\begin{equation}
V_{eff}=V_1 + V_2 + V_3 + SC,
\end{equation}
with
\begin{equation}
\begin{split}
V_1 &= -\fr{1}{12}M^{MN}f_{NP}{}^Lf_{ML}{}^P+M^{MN}f_{MP}{}^Lf_{LN}{}^P,\\
V_2 &= \fr12M^{MN}f_{PM}{}^Lf_{LN}{}^P,\\
V_3 &= M^{MN}M^{KL}M_{RS}\left(\fr{1}{12} f_{MK}{}^Rf_{NL}{}^S-\fr12f_{KM}{}^Rf_{NL}{}^S\right),\\
SC  &= \fr12Y^{MN}_{KL}f_{MR}{}^Kf_{NS}{}^L M^{RS}.
\end{split}
\end{equation}
Again the part $V_2$ is the full derivative and can be dropped.

It is straightforward to check the following identities
\begin{equation}
\label{YXE}
\begin{split}
Y^{RL}_{SM}X_{KR}{}^S&=-5X_{KM}{}^L,\\
Y^{RL}_{SM}f_{KR}{}^S&=-5f_{KM}{}^L,\\
Y^{KA}_{BL}X_{AN}{}^B&=X_{LN}{}^K+4X_{NL}{}^K,
\end{split}
\end{equation} 
that can be derived exactly in the same fashion as \eqref{YXSO}. The analogue of the second line of \eqref{Gm} is
\beq
\label{Md}
M_{\br{M}{N}}d^{\bar{N}\br{K}{L}}=d_{\bar{M}\br{N}{R}}M^{\br{N}{K}}M^{\br{R}{L}}
\eeq
and implies that the indices of the invariant tensor are raised and lowered by the generalised metric. This is in agreement with the definition of the unimodular matrix
\begin{equation}
M_{MN}=\mc{V}_M{}^{ij}\mc{V}_N{}^{kl}\W_{ik}\W_{jl}
\end{equation}
and the following representation of the invariant tensor \cite{Bergshoeff:2007ef}
\beq
d_{MNK}=\mc{V}_M{}^{ij}\mc{V}_N{}^{kl}\mc{V}_K{}^{mn}\W_{jk}\W_{lm}\W_{ni}
\eeq
if one takes into account the condition $\mc{V}_M{}^{ij}\W_{ij}=0$.

Using the identities \eqref{YXE}, the definition \eqref{Md} and the last line of \eqref{rel} we deduce for the effective potential
\begin{equation}
\label{VEt}
\begin{split}
V_{eff}=&-\fr{1}{12}X_{MK}{}^LX_{NL}{}^KM^{MN}+\fr{1}{12}X_{MR}{}^KX_{NS}{}^LM^{MN}M^{RS}M_{KL}+\\
&+\fr{1}{10}X_{RM}{}^KX_{NS}{}^LM^{MN}M^{RS}M_{KL}.
\end{split}
\end{equation}
The first term can be verified using the same technique as in the previous section. Namely, substituting the structure constant $X_{MN}{}^K$ from \eqref{X2f} and taking into account the identities \eqref{YXE} one obtains that the first term in the equation above is $V_1+SC$ plus a full derivative term.

The derivation of the second and the third term is longer but straightforward. Lets sketch the idea here on the example of the second term $X_{MR}{}^KX_{NS}{}^LM^{MN}M^{RS}M_{KL}$. Substituting here the expression \eqref{X2f} and expanding the brackets one obtains terms of the types
\begin{equation}
\label{termsE}
\begin{aligned}
& f_{MR}{}^Kf_{NS}{}^LM^{MN}M^{RS}M_{KL}, & & f_{RM}{}^Kf_{NS}{}^LM^{MN}M^{RS}M_{KL} \\
& f_{MR}{}^KY^{LA}_{BN}f_{AS}{}^BM^{MN}M^{RS}M_{KL}, & & f_{RM}{}^KY^{LA}_{BN}f_{AS}{}^BM^{MN}M^{RS}M_{KL},\\
& Y^{KQ}_{MP}f_{QR}{}^PY^{LA}_{BN}f_{AS}{}^BM^{MN}M^{RS}M_{KL}.
\end{aligned}
\end{equation}
Lets show that the third term in the second line is exactly proportional to the second term in the first line. Substituting the invariant tensor $Y^{MN}_{KL}=10d^{MNP}d_{PKL}$ as using the relation \eqref{Md} two times one can verify the following identities
\begin{equation}
\begin{aligned}
 &f_{RM}{}^KY^{LA}_{BN}f_{AS}{}^BM^{MN}M^{RS}M_{KL} = 10  f_{RM}{}^Kd^{CLA}d_{CBN}f_{AS}{}^BM^{MN}M^{RS}M_{KL}=\\
 &10  f_{RM}{}^Kd_{CBN}f_{AS}{}^B d_{KPQ}M^{CP}M^{AQ}M^{MN}M^{RS}=\\
 &10  f_{RM}{}^Kf_{AS}{}^B d^{PMJ}M_{JB}d_{KPQ}M^{AQ}M^{RS}=f_{RM}{}^Kf_{AS}{}^B Y^{MJ}_{KQ}M_{AB}M^{AQ}M^{RS}=\\
 &-5f_{RQ}{}^Jf_{AS}{}^BM_{JB}M^{AQ}M^{RS},
\end{aligned}
\end{equation}
where the identity \eqref{YXE} was used in the last line. Using the same idea one simplifies the last line in \eqref{termsE}. Finally, the contributions like the first term in the second line of \eqref{termsE} coming from two last terms in \eqref{VEt} precisely cancel each other.

After long algebraic calculations it can be derived that the expression \eqref{VEt} is up to a prefactor equal to the scalar potential of maximal gauged supergravity in 5 dimensions
\begin{equation}
V_{eff}=\fr92|A_1^{ij}|^2-\fr12|A_{2}^{i,jkl}|^2=\fr32V_{scalar},
\end{equation}
where the $|\;|^2$ stands for the contraction of all indices. To show this one expresses the potential in terms of the $T$-tensor by making use the relation \eqref{XT}. Finally, rewriting the $T$-tensor in terms of the $A$-tensor as \eqref{TA} and using the properties of the $A$-tensors one obtains the scalar potential of the maximal gauged supergravity.\footnote{We found the computer algebra system Cadabra \cite{Cadabra, Peeters:2007wn} very useful in dealing with long calculations.} We refer the reader to Appendix B for the proof that the potential \eqref{VEt} is invariant under gauge transformations \eqref{G_transf}.

\section{Conclusions and Discussion}

The above results show that the idea of Scherk-Schwarz reduction works in detail for $d=5$ and 6 as well. Thus this work extends \cite{Berman:2012uy} where the case d=4 was done in detail and the details for other dimensions were only sketched. 

The most interesting point to mention here is that geometry of the extended space plays an important role in the picture presented above. It is not just a Kaluza-Klein reduction where fields does not depend on internal coordinates. The extended space should be an extended geometry analogue of a  parallelisable space so the dependence on the dual coordinates should be of a particular form. These constraints match the quadratic constraint on the embedding tensor of gauged supergravity. Although there are many papers \cite{Coimbra:2011ky, Hohm:2011si, Hohm:2012gk} considering geometry of the extended space it is not fully understood how to describe this object. In this work we investigate a very particular situation but we hope it may contribute to the full picture of the extended geometry.

\section{Acknowledgements}
The author would like to thank Henning Samtleben for discussions on gauged supergravities, David Berman, Axel Kleinshmidt and Daniel Thompson for general discussions on extended geometry and Scherk-Schwarz reductions. 

This work is supported by Queen Mary CDTA fellowship. 

\section{Appendix}
\subsection{$SO(5,5)$ gaugings}

While the trombone is obtained in a straightforward way from the gauge group generators $X_{MN}{}^K$ one has to do some algebra to get the remained gauging $\q^{iM}$. This section is to show how this gauging can be obtained by suitable projections of the gauge group generators.

The gauge group generators $X_{MN}{}^K$ evaluated in the representation $\mc{R}_V$ have the form
\beq
X_{MN}{}^K=\Q_M^\a (t_\a)_N{}^K+\left(\fr{16}{5} (t^\a)_M{}^P(t_\a)_N{}^K + \d_M^P\d_N^K\right)\q_{P},
\eeq
where $t_\a$ are the generators of the global duality group and are given by $(\G_{ij})_N{}^K$. The embedding tensor reads
\beq
\Q_M{}^{ij}=-\q^{L[i}\G^{j]}{}_{LM}.
\eeq
Thus the gaugings can be explicitly separated out by the following contractions
\beq
\label{XcontrSO}
X_{MN}{}^K(\G^i{}_j)_K{}^N\G^{jMR}=128\,\q^{iR}-\fr{144}{5}\G^{iRS}\q_S.
\eeq
By making use of the first line of the definitions \eqref{Gm} one can show that the generators \eqref{Xgen} contracted in the same way give exactly \eqref{XcontrSO} with gaugings defined as \eqref{gaugingsSO}. Indeed, lets rewrite the generators $X_{MN}{}^K$ using the second line of \eqref{rel}
\beq
X_{MN}{}^K=f_{MN}{}^K+\fr18(\G_{ij})_N{}^K(\G^{ij})_C{}^Bf_{BM}{}^C+\fr14\d_N^Kf_{BM}{}^B.
\eeq
Contracting with the generator and the gamma matrix as in \eqref{XcontrSO} we obtain
\beq
X_{MN}{}^K(\G^i{}_j)_K{}^N\G^{jMR}=(f_{MN}{}^K-4f_{NM}{}^K)(\G^i{}_j)_K{}^N\G^{jMR}
\eeq
To show that this is exactly \eqref{XcontrSO} one needs to do some simple algebra and use the following identities
\beq
\begin{split}
\q^{iM}&=-\fr14\G^{jMD}f_{Dj}{}^i-\fr25\G^{iMN}\q_N,\\
f_{AP}{}^R&=\fr14\G_{iPQ}\G^{jQR}f_{Aj}{}^i-\fr14f_{AB}{}^B\d_P^R,\\
f_{Aj}{}^i\G_{i}{}^{AB}\h^{jk}&=4\q^{kB}+\fr85\G^{kBC}\q_C+\fr14\G^{kAB}f_{AR}{}^R,\\
Y^{BK}_{CL}f_{AK}{}^L&=-3f_{AC}{}^B-2f_{KA}{}^K\d_C^B-8\d_C^B\q_A.
\end{split}
\eeq
Here the first line is just a rewriting of \eqref{gaugingsSO}, the second line is a consequence of the definition \eqref{Gm} and the last line here. Finally, the third and the last lines are obtained directly by making use of  properties of twist matrices.

\subsection{Effective potential for $SO(5,5)$ case}

Since the generalised metric $M_{MN}$ is a coset representative we identify it with the unimodular matrix of \cite{Bergshoeff:2007ef} that has the same meaning and is defined as
\begin{equation}
\label{M_V}
M_{MN}={V}_{M}{}^{\a\dot{\a}}{V}_{N}{}^{\b\dot{\b}}\W_{\a\b}\W_{\dot{\a}\dot{\b}},
\end{equation}
where $\W_{\a\b}$ and $\W_{\dot{\a}\dot{\b}}$ are the symplectic invariants of $Spin(4)$ corresponding to each $SO(5)$  in the coset. These matrices are antisymmetric $\W_{\a\b}=-\W_{\b\a}$ and are used to raise and lower spinor indices $\W_{\a\b}\W^{\b\m}=\d_\a{}^\m$. The matrices $V_M^{\a\dot{\a}}$ are coset representatives of 
\begin{equation}
\fr{SO(5,5)}{SO(5)\times SO(5)}.
\end{equation}

Recall the effective potential \eqref{VSOt} that comes from Scherk-Schwarz reduction of M-theory in the extended space formalism
\begin{equation}
\label{VSOt1}
V_{eff}=-\fr{1}{8}X_{MK}{}^LX_{NL}{}^KM^{MN}+\fr{1}{32}X_{MR}{}^KX_{NS}{}^LM^{MN}M^{RS}M_{KL}.
\end{equation}
To show that this expression exactly reproduces the scalar potential of $D=6$ gauged supergavity one needs the following relation
\begin{equation}
\label{Gamma_V}
\begin{aligned}
\G_{iAB}V^{B\b\dot{\b}}=\mc{V}_i^{\hat{a}}(\g^{\hat{a}})_\a{}^\b V_A^{\a\dot{\b}}-
\mc{V}_i^{\hat{\dot{a}}}(\g^{\hat{\dot{a}}})_{\dot{\a}}{}^{\dot{\b}} V_A^{\b\dot{\a}},
\end{aligned}
\end{equation}
that follows from the invariance of the $SO(5,5)$ gamma-matrices \cite{Bergshoeff:2007ef}.

Consider the first term of the potential since it is easier to proceed. The calculations for the second term are longer but the idea is the same. In the absence of the trombone gauging the structure constants read
\begin{equation}
X_{MN}{}^K=-\q^{iL}\G^j_{LM}(\G_{ij})_N{}^K.
\end{equation}
Taking into account the quadratic constraint $\q^{iM}\q^{jN}\h_{ij}=0$, where $\h_{ij}$ is 10-dimensional flat metric and simple gamma-matrix algebra, one can write
\begin{equation}
V_1=-\fr18X_{MK}{}^LX_{NL}{}^KM^{MN}=-2\,\q^{iA}\q^{kB}\G_{iBN}\G_{kAM}M^{MN}.
\end{equation}
The next step is to substitute the explicit expression of the generalised metric $M_{MN}$ in terms of the coset representatives \eqref{M_V} and use the identity \eqref{Gamma_V}. This gives
\begin{equation}
\begin{aligned}
V_1=-2\q^{iA}\q^{kB}&\left(\mc{V}_i^{\hat{a}}(\g^{\hat{a}})_\m{}^\a V_B^{\m\dot{\a}}-
\mc{V}_i^{\hat{\dot{a}}}(\g^{\hat{\dot{a}}})_{\dot{\m}}{}^{\dot{\a}} V_B^{\a\dot{\m}}\right)\times\\
&\left(\mc{V}_k^{\hat{b}}(\g^{\hat{b}})_\n{}^\b V_A^{\n\dot{\b}}-
\mc{V}_k^{\hat{\dot{b}}}(\g^{\hat{\dot{b}}})_{\dot{\n}}{}^{\dot{\b}} V_A^{\b\dot{\n}}\right)\W_{\a\b}\W_{\dot{\a}\dot{\b}}.
\end{aligned}
\end{equation}
Using the definition of the T-tensor \eqref{T_tensor} this expression can be written only in terms of $(T^{\hat{a}})_{\a\dot{\a}}$ and $(T^{\hat{\dot{a}}})_{\a\dot{\a}}$
\begin{equation}
\begin{aligned}
V_1=&2(T^{\hat{a}})_{\n\dot{\a}}(T^{\hat{b}})^{\m\dot{\a}}(\g^{\hat{a}})_{\m}{}^{{\a}}(\g^{\hat{b}})_{\a}{}^{{\n}}
-2(T^{\hat{\dot{a}}})_{\n\dot{\a}}(T^{\hat{b}})^{\a\dot{\m}}(\g^{\hat{\dot{a}}})_{\dot{\m}}{}^{\dot{\a}}(\g^{\hat{b}})_{\a}{}^{{\n}}-\\
&2(T^{\hat{{a}}})_{\a\dot{\n}}(T^{\hat{\dot{b}}})^{\m\dot{\a}}(\g^{\hat{{a}}})_{{\m}}{}^{{\a}} (\g^{\hat{\dot{b}}})_{\dot{\a}}{}^{\dot{\n}}+2
(T^{\hat{\dot{a}}})_{\a\dot{\n}}(T^{\hat{\dot{b}}})^{\a\dot{\m}}(\g^{\hat{\dot{a}}})_{\dot{\m}}{}^{\dot{\a}} (\g^{\hat{\dot{b}}})_{\dot{\a}}{}^{\dot{\n}},
\end{aligned}
\end{equation}
where one should not that the matrices $\g^{\hat{a}}$ and $\g^{\hat{\dot{a}}}$ are antisymmetric. Reversing the order of the gamma matrices in the first and the last terms one obtains
\begin{equation}
\begin{aligned}
V_1=&4(T^{\hat{a}})_{\a\dot{\a}}(T^{\hat{a}})^{\a\dot{\a}}-4T_{\a\dot{\a}}T^{\a\dot{\a}}+4(T^{\hat{a}})_{\a\dot{\a}}(T^{\hat{a}})^{\a\dot{\a}}\\
=&8\Tr\left[T^{\hat{a}}\tilde{T}^{\hat{a}}-\fr12T\tilde{T}\right],
\end{aligned}
\end{equation}
where the identities in the last line of \eqref{T_tensor} and \eqref{TT} were used and tilde denotes transposition that implies $\Tr\left[T\tilde{T}\right]\equiv T_{\a\dot{\a}}T^{\a\dot{\a}}$.

The same but longer calculation shows that the second term in the potential $V_2$ gives the same expression up to  prefactor
\begin{equation}
V_2=-2\Tr\left[T^{\hat{a}}\tilde{T}^{\hat{a}}-\fr12T\tilde{T}\right].
\end{equation}
Together $V_1$ and $V_2$ result in
\begin{equation}
V_{eff}=6\Tr\left[T^{\hat{a}}\tilde{T}^{\hat{a}}-\fr12T\tilde{T}\right]=6V_{scalar}.
\end{equation}

\subsection{Invariance of the potential}

In this section we show the details of the proof that the potentials \eqref{VSOt} and \eqref{VEt} are invariant under the gauge transformations \eqref{G_transf}. In the dynamical picture of the extended geometry the potential is invariant due to the section condition. In the Scherk-Schwarz reduction of the theory the invariance of the potential is assured by the quadratic constraint \eqref{closure2}.

The terms that contribute to the effective potentials in $d=5,6$ 
\beq
\label{terms}
\begin{split}
&X_{MK}{}^LX_{NL}{}^KM^{MN},\\
&X_{MR}{}^KX_{NS}{}^LM^{MN}M^{RS}M_{KL},\\
&X_{RM}{}^KX_{NS}{}^LM^{MN}M^{RS}M_{KL}
\end{split}
\eeq
are invariant separately. Start with the first term. Its transformation gives
\beq
\begin{split}
&\d_\S\left(X_{MK}{}^LX_{NL}{}^KM^{MN}\right)=2X_{MK}{}^LX_{NL}{}^KX_{RS}{}^M\S^RM^{SN}=\\
&=-2[X_R,X_S]_K{}^LX_{NL}{}^K\S^RM^{SN}=-4\Tr[X_R,X_S,X_N]\S^{[R}M^{S]N}=\\
&=-2X_{SN}{}^K\Tr[X_K,X_R]\S^RM^{SN}=-2X_{(SN)}{}^KX_{KP}{}^QX_{RQ}{}^{P}\S^RM^{SN}=0,
\end{split}
\eeq
where we used the closure constraint \eqref{closure2} in the first line and cyclic symmetry of the trace in the second line. The last step here exploits the condition $X_{(AB)}{}^CX_{CK}{}^L=0$. 

For the transformation of the second term we have
\beq
\label{Vtransf1}
\begin{split}
&\d_\S(X_{MR}{}^KX_{NS}{}^LM^{MN}M^{RS}M_{KL})=2X_{MR}{}^K X_{NS}{}^LX_{PQ}{}^M\S^P M^{QN}M^{RS}M_{KL}+\\
&+2X_{MR}{}^K X_{NS}{}^LX_{PQ}{}^M\S^P M^{MN}M^{QS}M_{KL}-2X_{MR}{}^K X_{NS}X_{PK}{}^Q\S^{P}M^{MN}M^{RS}M^{QL}.
\end{split}
\eeq
After relabelling the indices the last two terms can be recast in the following form
\beq
\begin{split}
&(X_{MR}{}^KX_{PQ}{}^R-X_{MQ}{}^RX_{PR}{}^K)X_{NS}{}^L\S^{P}M^{MN}M^{QS}M_{KL}=\\
&=\left(X_P X_M-X_MX_P\right)_Q{}^K\S^{P}M^{MN}M^{QS}M_{KL}=\\
&=-X_{PM}{}^RX_{RQ}{}^K\S^{P}M^{MN}M^{QS}M_{KL}.
\end{split}
\eeq
This is exactly the first term in \eqref{Vtransf1} but with the opposite sign. Thus the second term in \eqref{terms} is invariant under the gauge transformations. The proof of the invariance of the third term is exactly the same.

\bibliographystyle{JHEP}
\bibliography{emusaev}
\end{document}